\documentclass{svmult}
\usepackage{amsmath,amsfonts,url}
\usepackage{pstricks,pst-coil}
\usepackage{graphicx}       
\usepackage[bottom]{footmisc}
\newcommand{\Cx}{{\mathbb C}}
\newcommand{\Ir}{{\mathbb Z}}
\newcommand{\Rl}{{\mathbb R}}
\newcommand{\B}{{\mathcal B}}
\renewcommand{\H}{{\mathcal H}}
\newcommand{\V}{{\mathcal V}}
\def\idty{{\mathchoice {\mathrm{1\mskip-4mu l}} {\mathrm{1\mskip-4mu l}} %
{\mathrm{1\mskip-4.5mu l}} {\mathrm{1\mskip-5mu l}}}}

\newcommand{\spec}{\mathop{\rm spec}}
\newcommand{\specrad}{\mathop{\rm spec rad}}
\renewcommand{\vec}[1]{\boldsymbol{#1}}

\newcommand{\ket}[1]{\left\vert #1\right\rangle}

\newcommand{\be}{\begin{equation}}
\newcommand{\ee}{\end{equation}}
\newcommand{\bea}{\begin{eqnarray}}
\newcommand{\eea}{\end{eqnarray}}
\newcommand{\beann}{\begin{eqnarray*}}
\newcommand{\eeann}{\end{eqnarray*}}
\newcommand{\eq}[1]{(\ref{#1})}
\begin{document}
\renewcommand{\thefootnote}{\fnsymbol{footnote}}
\title*{Ordering of Energy Levels\\ in Heisenberg Models and 
Applications$^*$}
\titlerunning{Ordering of Energy Levels in Heisenberg Models and 
Applications}
\author{Bruno Nachtergaele\inst{1}\and
Shannon Starr\inst{2}}
\institute{Department of Mathematics, University of California, Davis, Davis, CA 95616-8366, USA.   \texttt{bxn@math.ucdavis.edu}
\and 
Department of Mathematics, University of California, Los Angeles, Box 951555,
Los Angeles, CA 90095-1555, USA. \texttt{sstarr@math.ucla.edu}}
%
%
\maketitle

\footnotetext[1]{\copyright\ 2005 by the authors. This paper may be reproduced, in its entirety, for non-commercial purposes.}

\abstract{In a recent paper \cite{NSS} we conjectured that for ferromagnetic Heisenberg models the smallest eigenvalues in the invariant subspaces of fixed total spin are monotone decreasing as a function of the total spin and called this property  {\em ferromagnetic ordering of energy levels} (FOEL).
We have proved this conjecture for the Heisenberg model with arbitrary spins and coupling constants
on a chain \cite{NSS,NS_flm}.  In this paper we give a pedagogical introduction to this result and also
discuss some extensions and implications. The latter include the property that the relaxation time of symmetric simple exclusion processes on a graph for which FOEL can be proved, equals the relaxation time of a random walk on the same graph. This equality of relaxation times is known as Aldous' Conjecture.}

\section{Introduction}
\label{sec:intro}

The ferromagnetic Heisenberg model is the primordial quantum spin model. It has been studied
almost continuously since it was introduced by Heisenberg in1926. In the course of its long history,
this model has inspired an amazing variety of new developments in both mathematics and physics. The Heisenberg Hamiltonian is one of the basic, non-trivial quantum many-body operators,
and understanding its spectrum has been a guiding problem of mathematical physics for generations.

A lot of attention has been given to the Bethe-Ansatz solvable one-dimensional spin-$1/2$ model,
which has an infinite-dimensional algebra of symmetries \cite{JM}. The results we will discuss here
are not related to exact solutions but there is an essential connection with the $SU(2)$ symmetry
of the model, much in the spirit of the famous result by Lieb and Mattis (\cite{LM},
see also \cite[footnote 6]{Lieb}). The Lieb-Mattis Theorem proves ``ordering of energy levels'' for a 
large class of antiferromagnetic Heisenberg models on bipartite lattices.
Namely, if the two sublattices are $A$ and $B$, and all
interactions within $A$ and $B$ are ferromagnetic
while interactions in between $A$ and $B$ are antiferromagnetic,
then the unique ground state multiplet has
total spin equal to $|\mathcal{S}_A - \mathcal{S}_B|$,
where $\mathcal{S}_A$ and $\mathcal{S}_B$ are the
maximum total spins on the two sublattices.
Moreover, the minimum energy in the invariant subspace
of total spin $S$, for $S\geq |\mathcal{S}_A - \mathcal{S}_B|$,
is monotone increasing as a function of $S$. The most important example 
where this theorem provides useful information is the usual antiferromagnet
on a bipartite lattice with equal-size sublattices. Then the ground state is a unique spin singlet,
and the minimum energy levels for each possible total spin $S$, are
monotone increasing in $S$. Our aim is a similar result for ferromagnets. To be able 
to state the ferromagnetic ordering of energy levels (FOEL) property precisely, we
first give some definitions.

Let $\Lambda$ be a finite connected graph with a set of vertices or sites, $x$, that we will also 
denote by $\Lambda$ and a set $E$ of unoriented edges, or bonds, $(xy)$. We will often write 
$x\sim y \in\Lambda$ to signify that the edge $(xy)$ is present in $\Lambda$. In many physical 
examples one has $\Lambda \subset \Ir^d$.

Each site $x\in\Lambda$ has a quantum spin of  magnitude $s_x\in\{1/2,1,3/2,\ldots\}$, associated with it. The state space at $x$ is $2s_x+1$-dimensional and we denote by $S^i_x$, $i=1,2,3$, the standard spin-$s_x$ matrices acting on the $x$th tensor factor in the Hilbert space $\H=\bigotimes_{x\in\Lambda}
\Cx^{2s_x+1}$. The isotropic (also called XXX) ferromagnetic Heisenberg Hamiltonian on $\Lambda$ 
is given by
\be
H_\Lambda= - \sum_{x\sim y\in\Lambda} J_{xy} \vec{S}_x\cdot\vec{S}_y,
\label{heisen}\ee
where the real numbers $J_{xy}$ are the coupling constants, which we will always
assume to be strictly positive (that they are positive is what it means to have the {\em ferromagnetic} Heisenberg model). This model is widely used to describe ferromagnetism at the microscopic level whenever itinerant electron effects can be ignored. Examples are magnetic
domain walls and their properties and a variety of dynamical phenomena.

The spin matrices generate an irreducible representation of $SU(2)$ at each vertex. This
representation is conventionally denoted by $D^{(s_x)}$. An important feature of the Hamiltonian
\eq{heisen} is that it commutes with $SU(2)$ via the representation 
\be
\bigotimes_{x\in\Lambda} D^{(s_x)}
\label{tensorrep}\ee
or, equivalently, with the total spin matrices defined by
$$
S^i_\Lambda=\sum_{x\in\Lambda} S^i_x,\quad i=1,2,3.
$$
and hence also with the Casimir operator given by
$$
C=\vec{S}_\Lambda\cdot\vec{S}_\Lambda.
$$
The eigenvalues of $C$ are $S(S+1)$, $S= S_{\rm min}, S_{\rm min}+1,\ldots,S_{\rm max}
\equiv\sum_{x\in\Lambda} s_x$, which are the spin labels of the irreducible representations that occur in the direct sum decomposition of the tensor product representation \eq{tensorrep} into irreducible
components. The value of $S_{\rm min}$ is usually $0$ or $1/2$, but may be larger if one of the
$s_x$ is greater than $S_{\rm max}/2$. The decomposition into irreducible components can be 
obtained by repeated application of the Clebsch-Gordan series:
\be
D^{(s_1)}\otimes D^{(s_2)}\cong D^{(\vert s_1-s_2\vert)}
\oplus D^{(\vert s_1-s_2\vert + 1)}\cdots \oplus D^{(s_1 + s_2)}.
\label{CG}\ee
The label $S$ is called the {\em total spin}, and the eigenvectors of the eigenvalue $S(S+1)$
of $C$, are said to have total spin $S$. Let $\H^{(S)}$ denote the corresponding eigenspace.
Since $C$ commutes with $H_\Lambda$, the spaces $\H^{(S)}$ are invariant subspaces
for $H_\Lambda$. For any hermitian matrix $H$ leaving the spaces $\H^{(S)}$ invariant we define
$$
E(H,S)=\min\spec H \vert_{\H^{(S)}}.
$$

By Ferromagnetic Ordering of Energy Levels (FOEL) we mean the property 
$$
E(H,S)< E(H,S^\prime), \text{ if } S^\prime < S.
$$
for all $S$ and $S^\prime$ in the range $[S_{\rm min},S_{\rm max}]$.

In particular, if $H_\Lambda$ has the FOEL property it follows that its ground state energy is $E(H_\Lambda,S_{\rm max})$, which is indeed well-known to be the case for the Heisenberg ferromagnets. Moreover, since the multiplet of maximal spin is unique, FOEL also implies 
that the gap above the ground state is $E(H_\Lambda, S_{\rm max}-1)-E(H_\Lambda, S_{\rm max})$, which is well-known for translation invariant Heisenberg ferromagnets on Euclidean lattices.

\begin{conjecture}\label{con:foel}
All ferromagnetic Heisenberg models have the FOEL property.
\end{conjecture}

\begin{figure}\centering
\resizebox{!}{9truecm}{\includegraphics{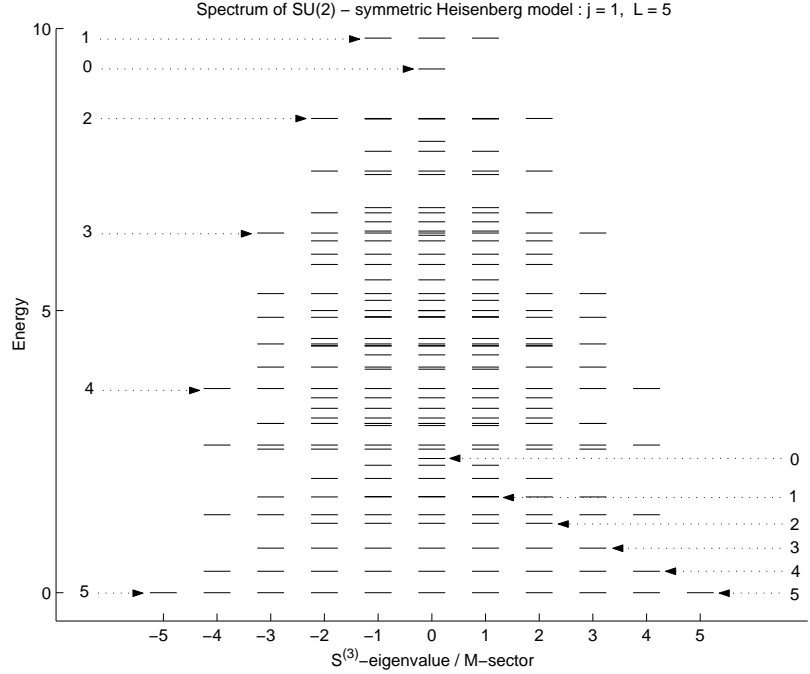}} 
\caption{\label{fig:spectrum}The spectrum of a ferromagnetic Heisenberg chain 
consisting of $5$ spin-$1$ spins, and with constant couplings. On the horizontal axis
we have plottted the eigenvalue of the third component of the total spin.
The spectrum is off-set so that the ground state energy vanishes. The arrows on the right, with 
label $S$, indicate the multiplets of eigenvalues $E(H,S)$, i.e., the smallest eigenvalue
in the subspace of total spin $S$. The monotone ordering of the spin labels is the
FOEL property. On the left, we have indicated the largest eigenvalues
for each value of the total spin. The monotone ordering of their labels in the range $1,\ldots,
5$, is the content of the Lieb-Mattis theorem applied to this system.}
\end{figure}

The FOEL property and the Lieb-Mattis theorem applied to a spin-$1$ chain of $5$ sites is illustrated in Figure \ref{fig:spectrum}.

Our main result is a proof of this conjecture for the special case of arbitrary ferromagnetic
Heisenberg models on chains, i.e., one-dimensional model \cite{NSS, NS_flm, NS_inprep}.

\begin{theorem}\label{thm:main}
FOEL holds for ferromagnetic XXX spin chains, i.e., for all
\be
H= - \sum_{x=1}^{L-1}J_{x,x+1}(\frac{1}{s_x s_{x+1}} \vec{S}_x\cdot \vec{S}_{x+1} -1),
\label{chainham}\ee
for any choice of $s_x\in\{1/2,1,3/2,\ldots\}$ and $J_{x,x+1} >0$.
\end{theorem}

\section{Proof of the Main Result}

Our proof of (Theorem \ref{thm:main}) proceeds by a finite induction argument for a sequence of models with Hamiltonians $H_k=H_k^*$, $1\leq k \leq N$, on Hilbert spaces $\H_k$, with the following
properties:

(i) There is a unitary representation of $SU(2)$, $U_k$, on $\H_k$, that commutes with $H_k$.

(ii) There are isometries $V_k:\H_{k+1}\to \H_k\otimes\Cx^2$, interwining the representations 
$U_{k+1}$ and $U_k\otimes D^{(1/2)}$, i.e., $V_k U_{k+1}(g)=(U_k(g)\otimes D^{(1/2)}(g)) V_k$, for
all $g\in SU(2)$, and such that
$$
H_{k+1}\geq V_k^* (H_k\otimes\idty) V_k
$$

(iii)  $H_1$ has the FOEL property.

(iv) For every $S=0,1/2,1,3/2,\ldots$, for which $\H^{(S)}\neq\{0\}$, we have
$$
E(H_{k+1},S+1/2)\leq E(H_k,S).
$$

We will first present the induction argument using the assumptions (i)--(iv), and then construct the sequence $H_k$ satisfying these four assumptions. This argument is a generalization of results in
\cite{KN} and \cite{NSS}. The sequence of Hamiltonians will, roughly speaking, be a sequence of
systems of increasing size, starting with the trivial system of a single spin. Property (i) simply means that all models will have isotropic interactions. Property (ii) will closely guide the construction of our
sequence. Property (iii) will be trivial in practice, since $H_1$ will be a multiple of the identity on $\H_1$
in our applications. Property (iv) has a nice physical interpretation at least in some of the examples
we will consider (see Section \ref{sec:droplet}). It is our (in)ability to prove (iv) that limits the range of models for which we can prove FOEL.

\begin{theorem}\label{thm:induction}
Let $(H_k)_{1\leq k\leq N}$, be a sequence of Hamiltonians satisfying properties (i)-(iv).
Then, for all $k$, $H_k$ has the FOEL property.
\end{theorem}
\begin{proof}
Since $H_1$ has the FOEL property by assumption, it is sufficient to prove the
induction step. Consider the following diagram: 
\newcommand{\negeq}{\raisebox{-5pt}{\rotatebox{45}{\mbox{$\geq$\ }}}}
\newcommand{\sgt}{\rotatebox{270}{\mbox{\!\!$>$\ }}}
 $$
\begin{array}{ccccccccccccccccccc}
	E(H_k,S)& >_1  &
	E(H_k,S+1)& >  &
	E(H_k,S+2)\\
	\sgt_2 & \negeq_3 & \sgt_2 &  \negeq & \sgt  \\
	E(H_{k+1},S+\frac{1}{2}) & >_4 &
	E(H_{k+1},S+\frac{3}{2}) & > &
	E(H_{k+1},S+\frac{5}{2})
\end{array}
$$
The inequality labeled $1$ is FOEL for $H_k$, and inequality $2$ is property (iv) assumed
in the theorem. We will prove inequality $3$ (using inequality $1$) and, combined with
inequality $2$ this implies inequality $4$, which is the induction step.

As before, we use superscripts to Hilbert spaces to denote their subspaces of fixed total
spin.  To prove inequality $3$, we start from the variational principle:
\beann
E(H_{k+1},S+1/2)&=&\inf_{\phi\in\H_{k+1}^{(S+1/2)},\Vert\phi\Vert=1}
\langle\phi,H_{k+1}\phi\rangle\\
&\geq&\inf_{\phi\in\H_{k+1}^{(S+1/2)},\Vert V_k\phi\Vert=1}
\langle\phi,V^*_k(H_k\otimes \idty_2)V_k\phi\rangle\\
&\geq&\inf_{\psi\in(\H_k\otimes\Cx^2)^{(S+1/2)},\Vert \psi\Vert=1}
\langle \psi, (H_k\otimes \idty_2)\psi\rangle
\eeann
The first inequality uses the fact that $V_k$ is an isometry and property (ii).
For the second inequality we enlarged the subspace over which the infimum is taken.
 
Now, we use the Clebsch-Gordan series \eq{CG} to see that
$(\H_k\otimes\Cx^2)^{(S+1/2)}\subset(\H_k^{(S)}\oplus\H_k^{(S+1)})\otimes\Cx^2$. Therefore
$$
E(H_{k+1},S+1/2)\geq \min \{E(H_{k},S),E(H_{k},S+1)\}=E(H_{k},S+1).
$$
Clearly, $H_k\otimes \idty_2$ restricted to $(\H_k^{(S)}\oplus\H_k^{(S+1)})\otimes\Cx^2$
has the same spectrum as $H_k$ restricted to $\H_k^{(S)}\oplus\H_k^{(S+1)}$.
The last equality then follows from inequality $1$, i.e., the induction hypothesis.
This concludes the proof of Theorem \ref{thm:induction}
 \end{proof}

For the proof of Theorem \ref{thm:main} we will apply Theorem \ref{thm:induction} to
the sequence $(H_k)_{1\leq k\leq N}$, with $N=2S_{\rm max}$, $H_1=0$, and
$H_N=H$, constructed as follows: for each $k=1,\ldots,N-1$, the model with
Hamiltonian $H_{k+1}$ is obtained from $H_k$ in one of two ways: either
a new spin 1/2 is added to right of the chain, or the magnitude of the rightmost
spin is increased by $1/2$. In both cases, $S_{\rm max}$ goes up by $1/2$ at
each step, hence $N=2\sum_{x=1}^L s_x$. Each $H_k$ is of the form \eq{chainham},
and we have written the interactions in such a way that the coupling constants
$J_{x,x+1}$ can be taken to be independent of $k$, although this is not crucial since all 
arguments work for any choice of positive coupling constants at each step. The parameters
that change with $k$ are thus $L$ and the set of spin magnitudes $(s_x)_{x=1}^L$.
To be explicit, the two possible ways of deriving $H_{k+1}$ from $H_k$ are
summarized in Table \ref{tab:cases}.

\begin{table}[b]\label{tab:cases}\begin{center}
\caption{Summary of the $k$-dependence of the sequence of models used in the proof by induction
of Theorem \ref{thm:induction}.}
\begin{tabular}{|c||c|c|}\hline
parameter & Case I & Case II \\ \hline\hline
$L$ & $L_{k+1}=L_k +1$ & $L_{k+1}=L_k$\\ \hline
$\{s_x\}$ & $s_{L_k +1}(k)=0, s_{L_k +1}(k+1)=1/2$ & $s_{L_k }(k+1)=s_{L_k} (k)+1/2$\\ \hline
$\H$ & $\H_{k+1}=\H_k\otimes\Cx^2$ & $\H_{k+1}=V(\H_k\otimes\Cx^2)$\\ \hline
\end{tabular}\end{center}
\end{table}

The Hamiltonians are of the form
\be
H_k= - \sum_{x=1}^{L_k-1}J_{x,x+1}\left(\frac{1}{s_x(k) s_{x+1}(k)} \vec{S}_x\cdot \vec{S}_{x+1} -1
\right),
\label{hamk}\ee
where $S_x^i$, $i=1,2,3$, are the $2s_x(k)+1$ dimensional spin matrices. To simplify the notation, 
the dependence on $k$ will often be omitted further on. 

We now have a uniquely defined sequence of Hamiltonians $(H_k)_{1\leq l\leq N}$, with $H_1 =0$
and $H_N=H$. Next, we proceed to proving the properties (i)-(iv). Property (i) is obvious by
construction. Property (iii) is trivial since $H_1=0$. To verify property (ii), we need to distinguish
the two cases for the relation between $H_k$ and $H_{k+1}$, as given in Table \ref{tab:cases}.

For Case I, $U_{k+1}=U_k\otimes D^{(1/2)}$ and we can take the identity map for $V$. Property (ii) follows from the positivity of the additional interaction term in $H_{k+1}$:
$$
H_{k+1}=H_k + J_{L_k,L_k+1}\left(\frac{1}{s_{L_k}\cdot(1/2)}\vec{S}_{L_k}\cdot\vec{S}_{L_k+1}-1\right).
$$

For Case II, we have $\H_k=\H_l\otimes \Cx^{2s_{L_k}+1}$ and $\H_{k+1}=\H_l\otimes \Cx^{2s_{L_{k+1}}+1}$, for some $l<k$, possibly $l=0,\H_0=\Cx$. Since $s_{L_{k+1}}=
s_{L_k}+1/2$, there is a (up to a phase) unique $SU(2)$ intertwining isometry $W:
\Cx^{2s_{L_{k+1}}}\to \Cx^{2s_{L_k}+1}\otimes \Cx^2$, namely the $W$ that identifies
the spin $s_{L_{k+1}}$ subrepresentation in $D^{(s_{L_k})}\otimes D^{(1/2)}$. 
From the intertwining property, the irreducibility of the spin representations, and the $SU(2)$ commutation relations one deduces that there is a constant $c$ such that
$$
W^*(S^i_{L_k}(k)\otimes\idty)W=cS^i_{L_k}(k+1), \quad i=1,2,3.
$$
The constant $c$ is most easily determined by calculating the left and right hand sides
on a highest weight vector (a simultaneous eigenvector of $C$ and $S^3$ with eigenvalues 
$S(S+1)$ and $S$, respectively). One finds
$$
c=\frac{s_{L_k}(k)}{s_{L_k}(k+1)}
$$
Now, take $V=\idty_{\H_l}\otimes W$. It is then straightforward to check that
$$
V^*\left(\frac{1}{s_{L_{k-1}} s_{L_k}(k)}\vec{S}_{L_{k-1}}\cdot\vec{S}_{L_k}\otimes\idty_2\right)V
=\frac{1}{s_{L_{k-1}} s_{L_k}(k+1)}\vec{S}_{L_{k-1}}\cdot\vec{S}_{L_k},
$$
where the spin matrices on the left hand side are of the magnitude determined
by $s_{L_{k-1}}$ and  $s_{L_k}(k)$ , while on the right hand side they are the
magnitudes of the spins are $s_{L_{k-1}}$ and $ s_{L_k}(k+1)$. 

To prove Property (iv), we start by observing that
$$
\spec(H_k\vert_{\H_k^{(S)}})=\spec(H_k\vert_{\V_k^{(S)}})
$$
where $\V_k^{(S)}$ is the subspace of $\H_k$ of all highest weight vectors
of weight $S$. This is an invariant subspace for $H_k$ and for every eigenvalue
of $H_k\vert_{\H_k^{(S)}}$ there is at least one eigenvector in $\V_k^{(S)}$.
Let $d(k,S)$ denote the dimension of $\V_k^{(S)}$.
 
Property (iv) will be obtained as a consequence of the following proposition 
and a version of the Perron-Frobenius Theorem.

\begin{proposition}\label{prop:mono}
We have $d(k+1,S+1/2)\geq d(k,S)$ and there are bases $\B_k^{(S)}$ for
$\V_k^{(S)}$ such that the matrices $A^{(k,S)}$ of $H_k\vert_{\V_k^{(S)}}$ with respect to these bases have the following properties:
\beann
A^{(k,S)}_{ij}\leq 0,&&\mbox{for } 1\leq i\neq j\leq d(k,S), 1\leq k\leq N\\
A^{(k+1,S+1/2)}_{ij}\leq A^{(k,S)}_{ij},&&\mbox{for } 1\leq i,j \leq d(k,S), 1\leq k\leq N-1.
\eeann
\end{proposition}

For reasons of pedagogy and length, we will give the complete proof of this proposition 
only for the spin $1/2$ chain. The proposition provides the assumptions needed to apply
a slightly extended Perron-Frobenius theorem (see, e.g., \cite{Wie}), which we state below. 

The standard Perron-Frobenius Theorem makes several statements about square matrices with all 
entries non-negative, which we will call a non-negative matrix for short.
Recall that a non-negative matrix $A$ is called irreducible if there exists an integer $n\geq 1$ such that the matrix elements of $A^n$ are all strictly positive. The standard results are
the following: (i) every non-negative matrix has a non-negative eigenvalue equal to its spectral radius (hence it has maximal absolute value among all eigenvalues), and there is a corresponding non-negative eigenvector (i.e., with all components non-negative); (ii) if $A$ is an irreducible non-negative matrix there is a unique eigenvalue with absolute value equal to the spectral radius of $A$,
which is strictly positive and has algebraic (and hence geometric) multiplicity $1$. Its corresponding
eigenvector can be chosen to have all strictly positive components.

If $A$ is a square matrix
$A$ with all off-diagonal matrix elements non-positive, we will call $A$ irreducible if there
exists a constant $c$ such that $c\idty -A$ is irreducible according to the previous definition.
From the standard Perron-Frobenius Theorem it immediately follows that 
the eigenvalue with smallest real part of an irreducible matrix in
the last sense is real, has algebraic (and hence geometric) multiplicity
$1$, and that the corresponding eigenvector can be chosen to have all
components strictly positive.
In the following, we will repeatedly use the information provided by the standard Perron-Frobenius Theorem as described above without further reference. Let $\specrad(A)$ denote the spectral radius
of a square matrix $A$.

\begin{lemma}\label{lem:PFcompare}
Let $A=(a_{ij})$ and $B=(b_{ij})$ be non-negative $n\times n$ matrices, and assume that
$a_{ij}\leq b_{ij}$, for all $1\leq i,j \leq n$. Then
\be
\specrad (A) \leq \specrad (B) .
\label{rArB1}\ee
If $B$ is irreducible and there is at least one pair 
$ij$ such that $a_{ij} < b_{ij}$, then
\be
\specrad (A) < \specrad (B) .
\label{rArB2}\ee
Since the spectral radii are also the eigenvalues of maximal absolute value, the same 
relations holds for these eigenvalues.
\end{lemma}

\begin{proof}
Let $r=\specrad (A)$. Then $A$ has a non-negative eigenvector, say $v$, with eigenvalue $r$. If $a_{ij}\leq b_{ij}$, for all $1\leq i,j \leq n$, it is clear that there is a non-negative vector $w$ such that
 \be
 Bv=rv+w.
\label{Bv} \ee
This relation implies that $\Vert B^k\Vert \geq r^k$, for all positive integers $k$ and, hence,
$\specrad (B)\geq r$. This proves \eq{rArB1}.

To prove \eq{rArB2} for irreducible $B$ such that $a_{ij}\leq b_{ij}$ for at least one pair of indices,
let $k$ be a positive integer such that $B^k$ is strictly positive. This implies that $B^k v$ has
all strictly positive components. From this it is easy to see that the non-negative $w$ such that
$$
B^kv=r^k v + w
$$
cannot be the zero vector. Therefore there is $z\in\Rl$ with all strictly positive components such that
$$
B^{k+1}v = r^{k+1} v + z.
$$
Since $z$ is strictly positive, there exists $\varepsilon >0$ such that,
$\epsilon v \leq z$ componentwise, and therefore we can find $\delta >0$ such that
$$
B^{k+1}v = (r+\delta)^{k+1}v +z^\prime,
$$
with $z^\prime$ non-negative. We conclude that $\specrad(B)\geq r+\delta > \specrad (A)$.
\end{proof}

Note that the argument that proves this lemma could also be used to give a lower bound
for the difference of the spectral radii. Since we do not need it, we will not pursue this
here. The next theorem is an extension of Lemma \ref{lem:PFcompare}.

\begin{theorem}\label{thm:PF}
Let $A=(a_{ij})$ and $B=(b_{ij})$ be two square matrices of size $n$ and $m$, respectively, with $n\leq m$, both with all off-diagonal matrix elements non-positive, and such that $b_{ij}\leq a_{ij}$, for $1\leq i,j\leq n$.
Then
\be
\inf\spec (B) \leq \inf\spec (A)
\label{PFW1}\ee
If $B$ is irreducible and either  (i) there exists at least one pair $ij$, $1\leq i , j\leq n$, 
such that  $b_{ij} <  a_{ij}$;  or (ii) $b_{ij}  < 0$, for at least one pair $ij$ with at least one of the 
indices $i$ or $j > n$, then
\be
\inf \spec (B) < \inf \spec (A).
\label{PFW2}\ee
\end{theorem}

\begin{proof}
Let $c\geq 0$ be a constant such that the matrices $A^\prime=(a^\prime_{ij})=c\idty_n - A$ and $B^\prime = (b^\prime_{ij}) = c\idty_m - B$ are non-negative. 
Define $A^{\prime\prime}$ to be the $m\times m$ matrix obtained by extending $A^\prime$ with zeros:
$$
(a^{\prime\prime}_{ij}) = A^{\prime\prime} = \begin{bmatrix}A^\prime& 0\\
0 & 0\end{bmatrix}.
$$
It is easy to see that $a^{\prime\prime}_{ij}\leq b^\prime_{ij}$, for $1\leq i,j\leq m$. Therefore,
we can apply Lemma \ref{lem:PFcompare} with $A^{\prime\prime}$ playing the role of $A$,
and $B^\prime$ playing the role of $B$.
Clearly, $\specrad(A^{\prime\prime})=c-\inf \spec(A)$ and $\specrad(B^\prime)=c-\inf \spec(B)$.
Therefore, this proves \eq{PFW1}.

Similarly, \eq{PFW2} follows from the additional assumptions and  \eq{rArB2}.
\end{proof}

{\em Proof of Theorem \ref{thm:main}:} The remaining point was to prove property (iv) needed
in the assumptions of Theorem \ref{thm:induction}. We use Proposition \ref{prop:mono}, which we
will prove in the next section, and apply Theorem \ref{thm:PF} with $A=A^{(k,S)}$ and
$B=A^{(k+1,S+1/2)}$. This completes the proof.

\section{The Temperley-Lieb basis. Proof of Proposition \ref{prop:mono}.}
  
In the proof of Theorem \ref{thm:main} in the previous section we used the matrix representation
of the Hamiltonians restricted to the highest weight spaces given by
Proposition \ref{prop:mono}. We now give the complete proof of that proposition
for the spin $1/2$ chain and sketch the proof in the general case.

The main issue is to find a basis of the highest weight spaces with the desired properties.
Fortunately for us, such a basis has already been constructed and we only need to show that
it indeed had the properties claimed in Proposition \ref{prop:mono}. For the spin $1/2$ chain 
we will use the Temperley-Lieb basis \cite{TL}, and for the general case its generalization to
arbitrary spin representations introduced by Frenkel and Khovanov \cite{FK}.

\subsection{The basis for spin $1/2$}

\begin{figure}[t]
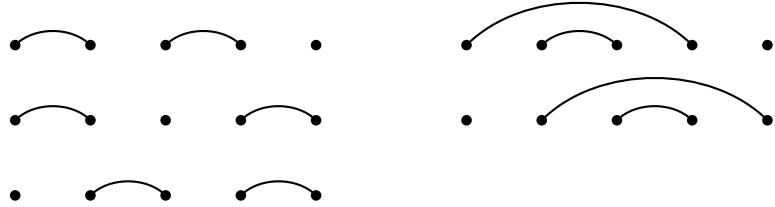
\centering
\pspicture(0,0)(21,5)
\psset{unit=.5cm,origin={0,0}}
\qdisk(1,5){2pt}\qdisk(3,5){2pt}\qdisk(5,5){2pt}\qdisk(7,5){2pt}\qdisk(9,5){2pt}
\psbezier(1,5)(1.5,5.5)(2.5,5.5)(3,5)\psbezier(5,5)(5.5,5.5)(6.5,5.5)(7,5)
\qdisk(13,5){2pt}\qdisk(15,5){2pt}\qdisk(17,5){2pt}\qdisk(19,5){2pt}\qdisk(21,5){2pt}
\psbezier(15,5)(15.5,5.5)(16.5,5.5)(17,5)\psbezier(13,5)(14.5,6.5)(17.5,6.5)(19,5)
\qdisk(1,3){2pt}\qdisk(3,3){2pt}\qdisk(5,3){2pt}\qdisk(7,3){2pt}\qdisk(9,3){2pt}
\psbezier(1,3)(1.5,3.5)(2.5,3.5)(3,3)\psbezier(7,3)(7.5,3.5)(8.5,3.5)(9,3)
\qdisk(13,3){2pt}\qdisk(15,3){2pt}\qdisk(17,3){2pt}\qdisk(19,3){2pt}\qdisk(21,3){2pt}
\psbezier(17,3)(17.5,3.5)(18.5,3.5)(19,3)\psbezier(15,3)(16.5,4.5)(19.5,4.5)(21,3)
\qdisk(1,1){2pt}\qdisk(3,1){2pt}\qdisk(5,1){2pt}\qdisk(7,1){2pt}\qdisk(9,1){2pt}
\psbezier(3,1)(3.5,1.5)(4.5,1.5)(5,1)\psbezier(7,1)(7.5,1.5)(8.5,1.5)(9,1)
\endpspicture
\caption{\label{fig:TL5-2} The possible configurations of $2$ arcs on $5$ vertices.}
\end{figure}

We start with the spin $1/2$ chain, i.e., $s_x=1/2$, for all $x$. In this case $S_{\rm max}=k/2$
and $V_k^{(S)}$ is the subspace of $(\Cx^2)^{\otimes k}$ consisting of all vectors $\psi$
such that $S^3\psi = S\psi$ and $S^+\psi =0$. Let $n$ be the ``spin-deviation''
defined as $S=k/2 - n$. Then, $n$ is a non-negative integer. The case  $n=0$ is trivial since 
$\dim \V_k^{(k/2)}=1$, namely just the mutiples of the vector $\ket{+}\otimes\ket{+}\otimes \cdots
\otimes\ket{+}$, where $\ket{\pm}$ is the basis of $\Cx^2$ that diagonalizes $S^3$.
For $n\geq 1$, the basis vectors are a tensor product of $n$ singlet vectors $\xi =
\ket{+}\otimes\ket{-}-\ket{-}\otimes\ket{+}$, accounting for two sites each, and $k-2n$ factors equal
to $\ket{+}$. Such vectors are sometimes called Hulth\'en brackets.
It is clear that any such factor is a highest weight vector of weight $k/2-n$, just
calculate the action of $S^3$ and $S^+$ on such a vector. They are not linearly independent
however, except in the trivial case $k=2$. The contribution of Temperley and Lieb was to show how to select a complete and linearly independent subset, i.e., a basis.
How to select the Temperley-Lieb basis, is most easily explained by representing the vectors
by configurations of $n$ arcs on the $k$ vertices $1,\dots,n$. The arcs are drawn above
the line of vertices as shown in Figure \ref{fig:TL5-2}. Each arc represents a spin singlet $\xi$,
and each unpaired vertex represents a factor $\ket{+}$. The vectors (configurations of arcs) selected for the basis are those that satisfy two properties: (i) the arcs are non-crossing, (ii) no arc spans an unpaired
vertex. The resulting set is a (non-orthogonal) basis. E.g., the basis for $k=5$ and $n=2$ is shown
in Figure \ref{fig:TL5-2}. We will use, $\alpha,\beta,\ldots$, to denote arc configurations that obey these rules, and by the corresponding basis vectors will be denoted by $\ket{\alpha},\ket{\beta},\ldots$.
We will use the notation $[xy]\in\alpha$ to denote that the arc connecting $x$ and $y$ is present
in $\alpha$.

{\em Proof of Proposition \ref{prop:mono} for the spin $1/2$ chain.}
The action of the Hamiltonian on the basis vectors has an appealing graphical
representation. We can write the Hamiltonian as
$$
H_k=-2\sum_{x=1}^{k-1} J_{x,x+1} U_{x,x+1}
$$
where $U_{x,x+1}=-\xi\otimes\xi^*$ which, up to a factor $-2$, is
the orthogonal projection
onto the singlet vector acting on the $x$th and $x+1$st factor in the tensor product.
The $U_{x,x+1}$ form a representation of the Temperley-Lieb algebra with parameter
$q=1$ (see, e.g,  \cite{KL}). It is a straightforward calculation to verify the action of $U_{x,x+1}$
on a basis vector $\ket{\alpha}$: (i) if both $x$ and $x+1$ are unpaired vertices in $\alpha$,
$U_{x,x+1}\ket{\alpha}=0$; (ii) if $[x, x+1]\in\alpha$, we have
$U_{x,x+1}\ket{\alpha}=-2\ket{\alpha}$; (iii) if $[uv]\in\alpha$, with exactly one of the vertices
$u$ and $v$ equal to $x$ or $x+1$, we have $U_{x,x+1}\ket{\alpha}=\ket{\beta}$,
where $\beta$ is obtained form $\alpha$ by removing $[uv]$ and adding $[x,x+1]$; (iv)
if $[ux]$ and $[x+1,v]$ are both present in $\alpha$, we have $U_{x,x+1}\ket{\alpha}=\ket{\beta}$,
where $\beta$ is obtained form $\alpha$ by removing $[ux]$ and $x+1,v]$, and adding
$[uv]$ and $[x,x+1]$. 

\begin{figure}
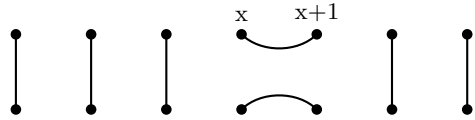
\centering
\pspicture(-2,-1)(12,1)
\psset{unit=.5cm,origin={0,0}}
\qdisk(0,0){2pt}\qdisk(2,0){2pt}\qdisk(4,0){2pt}\qdisk(6,0){2pt}\uput[90](6,0){x}
\qdisk(8,0){2pt}\uput[90](8,0){x+1}\qdisk(10,0){2pt}\qdisk(12,0){2pt}
\qdisk(0,-2){2pt}\qdisk(2,-2){2pt}\qdisk(4,-2){2pt}\qdisk(6,-2){2pt}
\qdisk(8,-2){2pt}\qdisk(10,-2){2pt}\qdisk(12,-2){2pt}
\psbezier(6,0)(6.5,-.5)(7.5,-.5)(8,0)\psbezier(6,-2)(6.5,-1.5)(7.5,-1.5)(8,-2)
\psline(0,0)(0,-2)\psline(2,0)(2,-2)\psline(4,0)(4,-2)\psline(10,0)(10,-2)\psline(12,0)(12,-2)
\endpspicture
\caption{\label{fig:Ux} The graphical representation of $U_{x,x+1}$.}
\end{figure}

\begin{figure}
\begin{center}
\resizebox{!}{3truecm}{\includegraphics{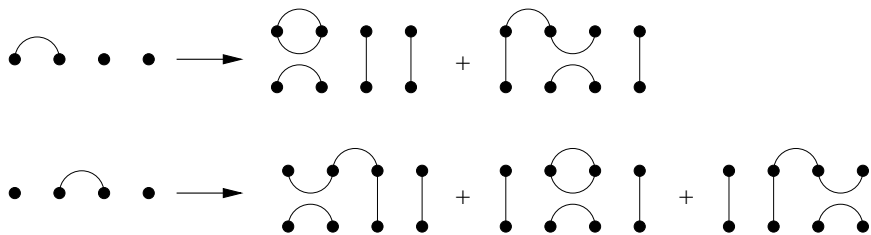}}\qquad
\end{center}
\caption{\label{fig:hulthen_brackets1}
Action of the Hamiltonian of the spin-1/2 XXX or XXZ
chain on a generalized Hulth\'en bracket, for $L=4$, $k=1$.}
\end{figure}

\begin{figure}[h]
\begin{center}
\resizebox{!}{5.5truecm}{\includegraphics{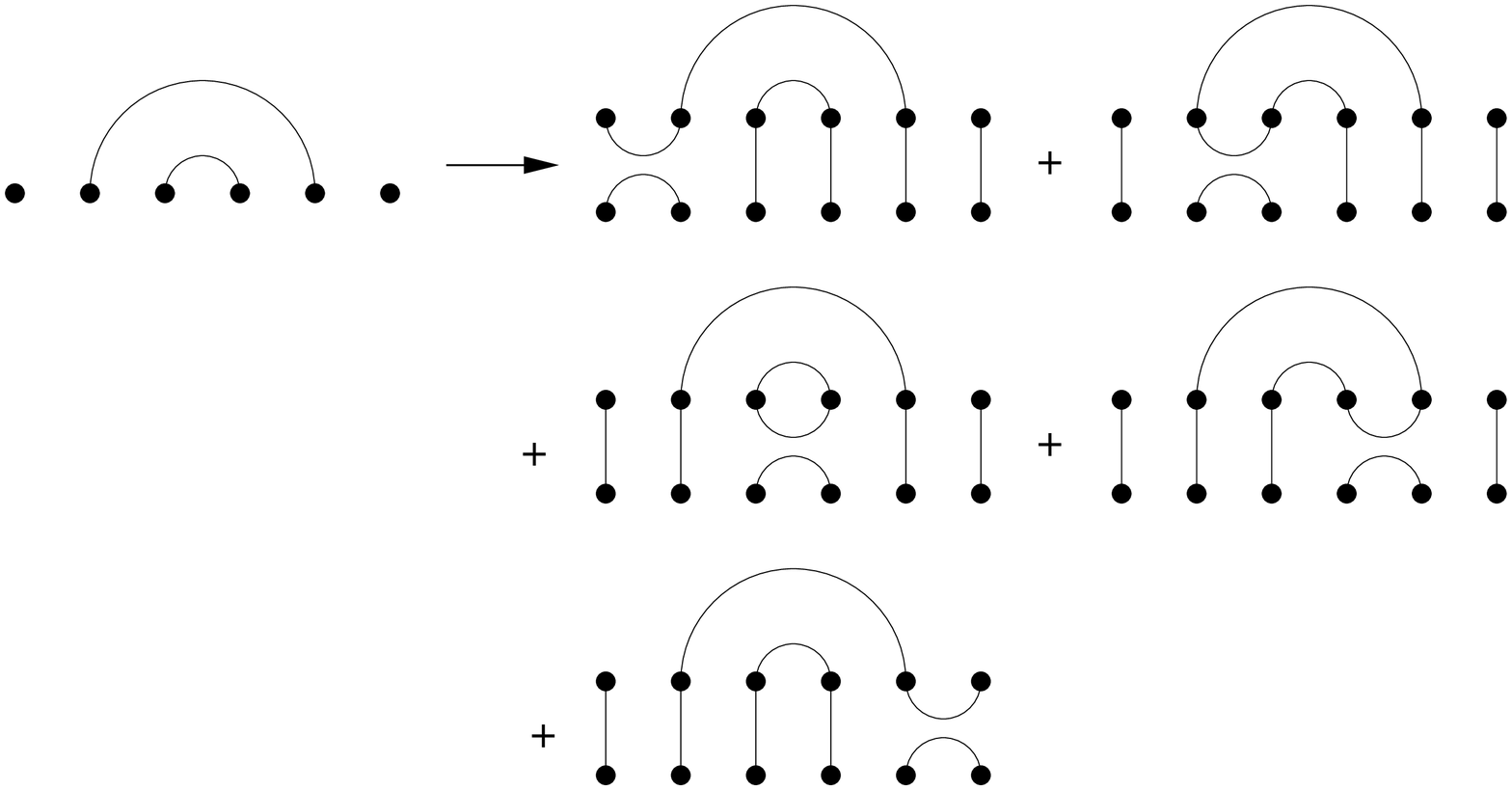}}\qquad
\end{center}
\caption{\label{fig:hulthen_brackets3}
Action of the Hamiltonian of the spin-1/2 XXX or XXZ
chain on a generalized Hulth\'en bracket, for $k=6$, $n=2$.}
\end{figure}

The action of $U_{x,x+1}$ on the vector $\ket{\alpha}$ can be graphically represented by 
placing the diagram shown in Figure \ref{fig:Ux} under the diagram for $\alpha$, and read
off the result using the graphical representation of the rules (i)--(iv) shown in Figure
\ref{fig:rules}. The action of the Hamiltonian is then obtained by summing over $x$
as shown in Figures \ref{fig:hulthen_brackets1} and \ref{fig:hulthen_brackets3} .

\newcommand{\TLone}{
\qdisk(0,0){2pt}\qdisk(2,0){2pt}\qdisk(0,2){2pt}\qdisk(2,2){2pt}
\psbezier(0,2)(.5,1.5)(1.5,1.5)(2,2)\psbezier(0,0)(.5,.5)(1.5,.5)(2,0)
}
\newcommand{\TLtwo}{
\qdisk(0,0){2pt}\qdisk(2,0){2pt}\qdisk(0,2){2pt}\qdisk(2,2){2pt}
\psbezier(0,2)(.5,1.5)(1.5,1.5)(2,2)\psbezier(0,0)(.5,.5)(1.5,.5)(2,0)
\psbezier(0,2)(.5,2.5)(1.5,2.5)(2,2)
\qdisk(5,1){2pt}\qdisk(7,1){2pt}\psbezier(5,1)(5.5,1.5)(6.5,1.5)(7,1)
}
\newcommand{\TLthree}{
\qdisk(0,0){2pt}\qdisk(0,2){2pt}\qdisk(2,0){2pt}\qdisk(4,0){2pt}\qdisk(2,2){2pt}\qdisk(4,2){2pt}
\psbezier(2,2)(2.5,1.5)(3.5,1.5)(4,2)\psbezier(2,0)(2.5,.5)(3.5,.5)(4,0)
\psbezier(0,2)(.5,2.5)(1.5,2.5)(2,2)\psline(0,0)(0,2)
\qdisk(7,1){2pt}\qdisk(9,1){2pt}\qdisk(11,1){2pt}\psbezier(9,1)(9.5,1.5)(10.5,1.5)(11,1)
}
\newcommand{\TLfour}{
\qdisk(0,0){2pt}\qdisk(0,2){2pt}\qdisk(2,0){2pt}\qdisk(4,0){2pt}\qdisk(2,2){2pt}\qdisk(4,2){2pt}
\psbezier(2,2)(2.5,1.5)(3.5,1.5)(4,2)\psbezier(2,0)(2.5,.5)(3.5,.5)(4,0)
\psbezier(0,2)(.5,2.5)(1.5,2.5)(2,2)\psline(0,0)(0,2)
\qdisk(6,0){2pt}\qdisk(6,2){2pt}
\psbezier(4,2)(4.5,2.5)(5.5,2.5)(6,2)\psline(6,0)(6,2)
\qdisk(9,1){2pt}\qdisk(11,1){2pt}\qdisk(13,1){2pt}\qdisk(15,1){2pt}
\psbezier(11,1)(11.5,1.5)(12.5,1.5)(13,1)
\psbezier(9,1)(10.5,2.5)(13.5,2.5)(15,1)
}
\begin{figure}
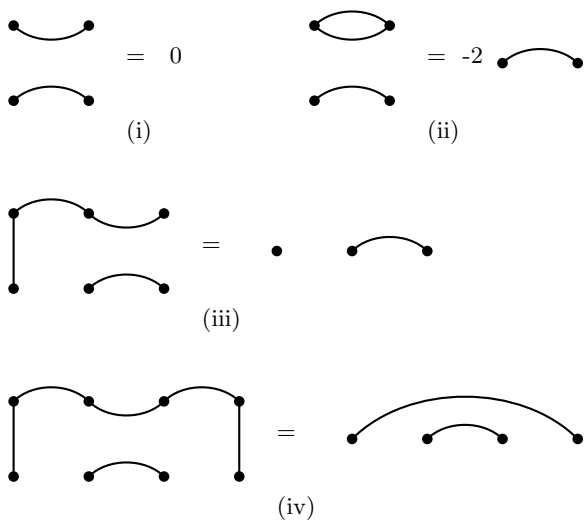
\centering
\pspicture(0,-6)(16,2)
\psset{unit=.5cm,origin={0,0}}
\put(3,1){=\quad 0}\put(3,-1){(i)}
\put(11,1){=\ \ -2}\put(11,-1){(ii)}
\put(5,-4){=}\put(5,-6){(iii)}
\put(7,-9){=}\put(7,-11){(iv)}
\TLone
\psset{origin={-8,0}}
\TLtwo
\psset{origin={0,5}}
\TLthree
\psset{origin={0,10}}
\TLfour
\endpspicture
\caption{\label{fig:rules} The graphical rules (i)-(iv) for the action of  $U_{x,x+1}$
on a Temperley-Lieb basis vector.}
\end{figure}

The important observation is the action of the Hamiltonian on a basis vector $\ket{\alpha}$ yields 
a linear combination of basis vectors with non-positive coefficients except possibly for
the coefficient of $\ket{\alpha}$ itself, which has the opposite
sign resulting from the ``bubble'' in the graphical representation. 
This means that all off-diagonal matrix elements are non-positive as claimed for the matrices
$A^{k,S}$ in the proposition.

The second will follow from the observation that $A^{k,S}$ is a submatrix
of $A^{k+1,S+1/2}$. Note that the spin deviation for $V_k^{(S)}$
and $V_{k+1}^{(S+1/2)}$ is the same, say $n$. Let us order the basis elements 
of $V_{k+1}^{(S+1/2)}$ so that all $\alpha$ where the last vertex, $k+1$, is unpaired,
are listed first, and consider the $\alpha\beta$ matrix element of $H_{k+1}$ for
such $\alpha$ and $\beta$. Then, it is easy to see that there are no contributions
from the $k,k+1$ term in Hamiltonian, since its action results in non-zero
coefficients only for configurations where $k+1$ belongs to an arc. This means
that these matrix elements are identical to those computed for $H_k$ for
basis vectors labeled $\alpha^\prime$ and $\beta^\prime$ obtained from
$\alpha$ and $\beta$ by dropping the last vertex, $k+1$ which is unpaired.

This completes the proof of Proposition \ref{prop:mono} in the case of the pure
spin $1/2$ chain. Q.E.D.
\subsection{The basis for higher spin}

We are looking for a basis of the space of highest weight vectors of weight $S$ of the
spin chain with Hilbert space $\H_k$. Equivalently, we may look for a basis of the
$SU(2)$ intertwiners $D^{(S)}\to \H_k$. There is a graphical algebra of such intertwiners 
with a very convenient basis, the dual canonical basis, introduced by Frenkel and Khovanov \cite{FK}.
This is the basis we will use, but we will present it as a basis for the subspaces $V_k^{(S)}$
of highest weight vectors.

The state space at site $x$ can be thought of as the symmetric part of $2s_x$
spins-$\frac{1}{2}$. We can label the $2s_x+1$ states by the Ising configurations
\newcommand{\da}{\downarrow}
\newcommand{\ua}{\uparrow}
$$
\ket{\ua\ua\cdots\ua},\, \ket{\da\ua\cdots\ua},\,
\ket{\da\da\ua\cdots\ua},\, \ldots,\, \ket{\da\da\da\cdots\da}\, .
$$
where each configuration stands for the equivalence class up to re-ordering of
all configurations with the same number of down spins. 
E.g., $\ket{\downarrow\downarrow\uparrow\uparrow\uparrow}$
is the vector normally labelled as $\ket{j,m} = \ket{5/2,1/2}$, and {\em not}
the tensor $\ket{\da}\otimes\ket{\da}\otimes\ket{\ua}\otimes\ket{\ua}\otimes\ket{\ua}$.
The states for a chain of $L$ spins of magnitudes $s_1,\ldots,s_L$
are then tensor products of these configurations. We shall call such
vectors {\em ordered Ising configurations}. These tensor product vectors, in
general, are not eigenvectors of the Casimir operator $S$, i.e., they are not 
of definite total spin. Suitable linear combinations that do have definite
total spin are obtained by extending the Hulth\'en bracket idea to
arbitrary spin as follows. Start from any ordered Ising configuration such that $2M=\#\ua -
\#\da$. Then, look for the leftmost $\da$   that has a $\ua$ to its left,
and draw an arc connecting this $\da$ to the rightmost $\ua$, left of it.
At this point, one may ignore the paired spins,
and repeat the procedure until there is no remaining unpaired $\da$ with
an unpaired $\ua$ to its left. This procedure guarantees that no arcs will
cross and no arc will span an unpaired spin. The result, when ignoring all paired
spins, is an ordered Ising  configuration of a single spin.
See Figure \ref{fig:FKbasis}
for an example of this procedure. 
The result is a basis for the spin chain consisting
entirely of simultaneous eigenvectors of the total spin
and its third component, with eigenvalues $S$ and $M$, respectively.
The value of $M$ is $1/2$ times the difference between the number
of up spins and the number of down spins in the ordered Ising
configuration.
The total-spin $S$ is equal
to $\mathcal{S}$ minus the number of pairs.  Clearly, the highest
weight vectors are then those that have no unpaired $\da$, i.e.,
the ordered Ising configuration consists exclusively of up spins.
\newcommand{\egpic}{
\begin{picture}(230,70)(0,30)
\put(10,80){\line(1,0){20}}  \put(10,80){\line(0,-1){10}}
\put(30,70){\line(-1,0){20}} \put(30,70){\line(0,1){10}}
\put(15,90){\vector(0,-1){10}}
\put(20,80){\vector(0,1){10}}
\put(25,80){\vector(0,1){10}}
\put(40,80){\line(1,0){15}}  \put(40,80){\line(0,-1){10}}
\put(55,70){\line(-1,0){15}} \put(55,70){\line(0,1){10}}
\put(45,90){\vector(0,-1){10}}
\put(50,80){\vector(0,1){10}}
\put(65,80){\line(1,0){25}} \put(65,80){\line(0,-1){10}}
\put(90,70){\line(-1,0){25}} \put(90,70){\line(0,1){10}}
\put(70,90){\vector(0,-1){10}}
\put(75,90){\vector(0,-1){10}}
\put(80,90){\vector(0,-1){10}}
\put(85,80){\vector(0,1){10}}
\put(100,80){\vector(1,0){10}}
\put(120,80){\line(1,0){20}} \put(120,80){\line(0,-1){10}}
\put(140,70){\line(-1,0){20}} \put(140,70){\line(0,1){10}}
\put(125,90){\vector(0,-1){10}}
\put(130,80){\vector(0,1){10}}
\put(145,80){\oval(20,20)[t]}
\put(150,80){\line(1,0){15}} \put(150,80){\line(0,-1){10}}
\put(165,70){\line(-1,0){15}} \put(165,70){\line(0,1){10}}
\put(160,80){\vector(0,1){10}}
\put(175,80){\line(1,0){25}} \put(175,80){\line(0,-1){10}}
\put(200,70){\line(-1,0){25}} \put(200,70){\line(0,1){10}}
\put(180,90){\vector(0,-1){10}}
\put(185,90){\vector(0,-1){10}}
\put(190,90){\vector(0,-1){10}}
\put(195,80){\vector(0,1){10}}
\put(210,80){\vector(1,0){10}}
\put(20,40){\line(1,0){20}} \put(20,40){\line(0,-1){10}}
\put(40,30){\line(-1,0){20}} \put(40,30){\line(0,1){10}}
\put(25,50){\vector(0,-1){10}}
\put(30,40){\vector(0,1){10}}
\put(45,40){\oval(20,20)[t]}
\put(50,40){\line(1,0){15}} \put(50,40){\line(0,-1){10}}
\put(65,30){\line(-1,0){15}} \put(65,30){\line(0,1){10}}
\put(70,40){\oval(20,20)[t]}
\put(75,40){\line(1,0){25}} \put(75,40){\line(0,-1){10}}
\put(100,30){\line(-1,0){25}} \put(100,30){\line(0,1){10}}
\put(85,50){\vector(0,-1){10}}
\put(90,50){\vector(0,-1){10}}
\put(95,40){\vector(0,1){10}}
\put(110,40){\vector(1,0){10}}
\put(130,40){\line(1,0){20}} \put(130,40){\line(0,-1){10}}
\put(150,30){\line(-1,0){20}} \put(150,30){\line(0,1){10}}
\put(135,50){\vector(0,-1){10}}
\put(167.5,40){\oval(55,35)[t]}
\put(155,40){\oval(20,20)[t]}
\put(160,40){\line(1,0){15}} \put(160,40){\line(0,-1){10}}
\put(175,30){\line(-1,0){15}} \put(175,30){\line(0,1){10}}
\put(180,40){\oval(20,20)[t]}
\put(185,40){\line(1,0){25}} \put(185,40){\line(0,-1){10}}
\put(210,30){\line(-1,0){25}} \put(210,30){\line(0,1){10}}
\put(200,50){\vector(0,-1){10}}
\put(205,40){\vector(0,1){10}}
\end{picture}}

\begin{figure}
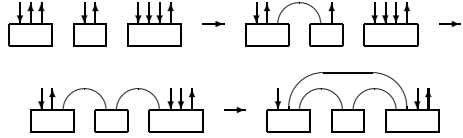
\centering
\resizebox{!}{2cm}{\egpic}
\caption{Construction of a basis vector from an ordered Ising
configuration}\label{fig:FKbasis}\end{figure}
The vectors can be expanded in the tensor product basis by the following procedure:
each arc is replaced by the spin singlet $\ket{\ua}\otimes\ket{\da}-\ket{\da}
\otimes \ket{\ua}$, and the unpaired spins are replaced by their tensor
products. Finally, one symmetrizes in each block.

Next, we briefly sketch how the properties claimed in Proposition \ref{prop:mono}
can be verified. To do this we have to calculate the action of the Hamiltonian on the
highest weight vector constructed in the previous paragraph. This is most easily 
accomplished by deriving a graphical representation for the action of each term in the
Hamiltonian as we did in the case of the pure spin $1/2$ chain.
The Heisenberg interaction for arbitrary spins of magnitude $s_x$ and $s_{x+1}$
can be realized as an interaction between spin $\frac{1}{2}$ 's making up the spin 
$s_x$ and $s_{x+1}$, conjugated with the projections onto the symmetric vectors.
The result is the following:
$$
-h_{x,x+1}
=\frac{1}{2}(\frac{1}{s_x s_{x+1}}\boldsymbol{S}_x \cdot \boldsymbol{S}_{x+1}-1)
= \begin{array}{c} \resizebox{1.85cm}{!}{\includegraphics{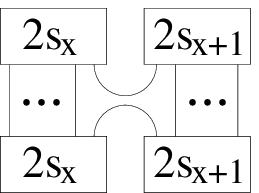}}
\end{array}\, .
$$
Here, the rectangles with label $2 s$ represent the symmetrizing projections
on the space of $2s$ spin $\frac{1}{2}$ variables.
The fundamental algebraic property
that allows us to calculate the matrix elements of $H_k$ graphically
is the Jones-Wenzl relation (c.f., \cite{KL} and references therein):
\newcommand{\jwonepic}{
\setlength{\unitlength}{789sp}%
\begin{picture}(3024,5724)(2389,-7473)
\thinlines
\put(2401,-3061){\framebox(1800,1200){}}
\put(4501,-3061){\framebox(900,1200){}}
\put(2901,-2686){$2s$}
\put(4726,-2686){$1$}
\put(2401,-7561){\framebox(1800,1200){}}
\put(4501,-7561){\framebox(900,1200){}}
\put(2901,-7136){$2s$}
\put(4726,-7136){$1$}
\put(2401,-5311){\framebox(3000,1200){}}
\put(2626,-6361){\line( 0, 1){1050}}
\put(2626,-4111){\line( 0, 1){1050}}
\put(3976,-4111){\line( 0, 1){1050}}
\put(3976,-6361){\line( 0, 1){1050}}
\put(4951,-6361){\line( 0, 1){1050}}
\put(4951,-4111){\line( 0, 1){1050}}
\put(2801,-4986){$2s+1$}
\put(2851,-5836){$\ldots$}
\put(2776,-3661){$\ldots$}
\end{picture}}

\newcommand{\jwtwopic}{
\setlength{\unitlength}{789sp}%
\begin{picture}(3024,5724)(2389,-7473)
\thinlines
\put(2401,-7561){\framebox(1800,1200){}}
\put(4551,-7561){\framebox(900,1200){}}
\put(2901,-7136){$2s$}
\put(4726,-7136){$1$}
\put(2401,-3061){\framebox(1800,1200){}}
\put(4551,-3061){\framebox(900,1200){}}
\put(2901,-2686){$2s$}
\put(4776,-2686){$1$}
\put(2401,-5311){\framebox(1800,1200){}}
\put(4551,-5311){\framebox(900,1200){}}
\put(2901,-4986){$2s$}
\put(4776,-4986){$1$}
\put(2626,-6361){\line( 0, 1){1050}}
\put(2626,-4111){\line( 0, 1){1050}}
\put(3976,-4111){\line( 0, 1){1050}}
\put(3976,-6361){\line( 0, 1){1050}}
\put(5001,-6361){\line( 0, 1){1050}}
\put(5001,-4111){\line( 0, 1){1050}}
\put(2851,-5836){$\ldots$}
\put(2776,-3661){$\ldots$}
\end{picture}}

\newcommand{\jwthreepic}{
\setlength{\unitlength}{789sp}
\begin{picture}(3024,5724)(2389,-7473)
\thinlines
\put(2401,-7561){\framebox(1800,1200){}}
\put(4501,-7561){\framebox(900,1200){}}
\put(2901,-7136){$2s$}
\put(4726,-7136){$1$}
\put(4551,-3061){\oval(1050,1050)[bl]}
\put(4551,-3061){\oval(1050,1050)[br]}
\put(4551,-6361){\oval(1050,1050)[tr]}
\put(4551,-6361){\oval(1050,1050)[tl]}
\put(2576,-6361){\line( 0, 1){1050}}
\put(2576,-4111){\line( 0, 1){1050}}
\put(3751,-4111){\line( 0, 1){1050}}
\put(3751,-6361){\line( 0, 1){1050}}
\put(2401,-3061){\framebox(1800,1200){}}
\put(4501,-3061){\framebox(900,1200){}}
\put(2401,-5311){\framebox(2600,1200){}}
\put(2700,-5836){$\ldots$}
\put(2700,-3661){$\ldots$}
\put(2600,-4886){$2s-1$}
\put(2901,-2686){$2s$}
\put(4726,-2686){$1$}
\end{picture}}
$$
\begin{array}{c} \jwonepic \end{array}\, =\, 
\begin{array}{c} \jwtwopic \end{array}\, +\, 
\frac{2s}{2s+1}\,
\begin{array}{c} \jwthreepic \end{array}\, 
$$

For any element of the basis introduced above one can now compute the action
of the Hamiltonian and write it as a linear combination of the same basis
vectors. From the grahical rules it is easy to observe that all off-diagonal matrix
elements are non-positive. 

As before, it is straighforward to identify the basis for $\V_k^{(S)}$ with a subset
of the basis for $\V_{k+1}^{(S+1/2)}$. The label of the rightmost box in any basis vector for the
system $k$ is raised by one but the number of arcs remains unchanged.

The crucial property that allows us to compare the two Hamiltonians is the
following. When $H_{k+1}$ acts on a basis vector obtained from a corresponding
$H_k$ vector as we have just described, the only possible new terms that are generated
are off-diagonal terms, which do not contain a bubble and, hence, are negative.
The details of the calculation of these matrix elements and further applications will appear
elsewhere \cite{NS_inprep}.

\section{Extensions}

A highly desirable extension of our main result, of course, would be the proof of 
Conjecture \ref{con:foel} for all ferromagnetic isotropic Heisenberg models on 
an arbitrary graph!  While we have been able to prove some partial results for the spin-$1/2$
model on an arbitrary tree and a few other graphs, we do not have an argument
that works for arbitrary graphs \cite{NS_inprep}. But there are a few other directions
in which one might extend the ordering of energy levels property. The aim of this
section is to discus two such generalizations. In the first, the group $SU(2)$ is replaced
by the quantum group $SU_q(2)$,  $0<q<1$. This only works on the chain and, as
far as we are aware, leads to information about physically interesting models in the
case of the spin $1/2$ chain, namely the XXZ chain. The second generalization we 
consider is isotropic ferromagnetic models with higher order nearest neighbor interaction
terms, such as $(\vec{S}_x\cdot\vec{S}_{x+1})^2$. For this to be relevant, the spins
have to be of magnitude $\geq 1$. 

\subsection{The spin 1/2 $SU_q(2)$-symmetric XXZ chain}

It is well-known that the translation invariant spin-$1/2$ XXZ chain with a particular
choice of boundary fields is $SU_q(2)$ invariant \cite{PS}.  This $SU_q(2)$ symmetry
can be exploited in much the same way as the $SU(2)$ symmetry of the isotropic model
\cite{KN,NS_droplet}. Here we will show how it leads to a natural $SU_q(2)$ analogue of the
FOEL property.

The Hamiltonian of the $SU_q(2)$-invariant ferromagnetic spin-$1/2$ chain of length $L\geq 2$ is
given by 
\bea
H_L&=&-\sum_{x=1}^{L-1}  [\Delta^{-1} (S^1_x S^1_{x+1} + 
S^2_x S^2_{x+1})+ (S_x^{3} S_{x+1}^{3} - 1/4)]\label{hamXXZ}\\
&&-A(\Delta)( S_L^3-S^3_1).\nonumber
\eea
where $\Delta >1$, and
$$
A(\Delta)=\frac{1}{2}\sqrt{1-1/\Delta^2}
$$
This model commutes with one of the two natural representation of $SU_q(2)$ on $(\Cx^2)^L$,
with $q\in (0,1)$, such that $\Delta=(q+q^{-1})/2$. Concretely, this means that $H_L$
commutes with the three generators of this representation defined as follows:
\beann
S^3\!&=&\!\sum_{x=1}^L \idty_1\otimes\cdots\otimes
S^3_x\otimes\idty_{x+1}\otimes\cdots\idty_L\label{spinmua}\\
S^+\!&=&\!\sum_{x=1}^L t_1\otimes\cdots\otimes t_{x-1}\otimes
S^+_x\otimes\idty_{x+1}\otimes\cdots\idty_L\label{spinmub}\\
S^-\!&=&\!\sum_{x=1}^L \idty_1\otimes\cdots\otimes
S^-_x\otimes t^{-1}_{x+1}\otimes\cdots t^{-1}_L
\eeann
where 
$$
t=q^{-2S^3}=\left(\begin{array}{cc} q^{-1}&0\\0&q\end{array}\right).
$$
The $SU_q(2)$ commutation relations are
$$
[S^3,S^\pm]=\pm S^\pm, \quad [S^+,S^-]=\frac{q^{2S^3}-q^{-2S^3}}{q-q^{-1}} \, .
$$
Note that one recovers the $SU(2)$ definitions and commutation relations
in the limit $q\to 1$. $H_L$ also commutes with the Casimir opeator for $SU_q(2)$, given by 
$$
C= S^+S^- + \frac{ (qT)^{-1}+qT}{(q^{-1}-q)^2}, \quad T= t\otimes t\otimes\cdots\otimes t.
$$
The eigenvalues of $C$ are 
$$
 \frac{q^{-(2S+1)}+q^{2S+1}}{(q^{-1}-q)^2}, \quad S=0,1/2,1,3/2,\ldots
$$
and play the same role as $S$ for the XXX model, e.g., they label the irreducible
representations of $SU_q(2)$. The eigenspaces of $C$ are invariant subspaces of
$H_L$ and, as before, we denote the smallest eigenvalues of $H_L$ restricted to
these invariant subspaces by $E(H_L,S)$. Note that the subspaces depend on $q$,
but their dimensions are constant for $0<q\leq 1$.
 
\begin{theorem}
$$
E(H_L, S+1) <  E(H_L,S), \quad\mbox{ for all } S\leq L/2-1.
$$
\end{theorem}

The proof of this theorem is identical to the one for the isotropic spin-$1/2$ chain
up to substitution of the singlet vector $\xi$ by the $SU_q(2)$ singlet $\xi_q
= q\ket{+}\otimes\ket{-}-\ket{-}\otimes\ket{+}$, and changing the scalar value of the
``bubble" to $-(q+q^{-1})$. The details are given in \cite{NSS}. 

\subsection{Higher order interactions}

For spins of magnitude greater than $1/2$ the Heisenberg interaction is not the only 
$SU(2)$ invariant nearest neighbor interactions. It is easy to show that 
the most general $SU(2)$ invariant interaction of two spins of magnitudes $s_1$
and $s_2$, i.e., any hermitian matrix commuting with the representation $D^{(s_1)}\otimes
D^{(s_2)}$, is an arbitrary polynomial of degree $\leq 2\min\{s_1,s_2\}$ in the Heisenberg 
interaction with real coefficients:
\be
h_{12}=\sum_{m=0}^{2\min\{s_1,s_2\}}J^{(m)}(\vec{S}_1\cdot\vec{S}_2)^m.
\label{h_general}\ee
The definition of the FOEL property only uses $SU(2)$-invariance and therefore
applies directly to any Hamiltonian for a quantum spin system on a graph 
with at each edge an interactionof the form \eq{h_general}.
We believe it is possible to determine the exact range of coupling constants 
$J^{(m)}$ such that FOEL holds for spin $s$ chains with translation invariant
interactions. So far, we have carried this out only for the spin-$1$ chain.

\begin{theorem}
FOEL holds for the spin-$1$ chains with Hamiltonian
$$
H_L=\sum_{x=1}^{L-1} (1-\vec{S}_x \cdot \vec{S}_{x+1}) + \beta(1-\vec{S}_x \cdot \vec{S}_{x+1})^2)
$$
with $0\leq \beta\leq 1/3$. Level crossings occur at $\beta =1/3$ and FOEL does not hold, in general, for
$\beta > 1/3$.
\end{theorem}

The overall method of proof is the same as for the standard Heisenberg model. 
Theorem \ref{thm:induction} applies directly, since only the $SU(2)$ symmetry is used
in its proof. The only difference is in the proof of Proposition \ref{prop:mono}. The same 
basis for the highest weight spaces is used but verifying the signs of the matrix elements is 
more involved.

\section{Applications}

In this section we discuss a number of results that are either consequences of the FOEL
property, or other applications of the properties of the Heisenberg Hamiltonian that allowed
us to prove FOEL.

\subsection{Diagonalization at low energy}

The most direct applications of the FOEL property are its implications for the low-lying
spectrum of the Hamiltonian. FOEL with strict inequality implies that the ground states
are the multiplet of maximal spin which, of course, is not a new result. Since the maximal
spin multiplet is unique, the first excited state must belong to less than maximal spin
and therefore, by FOEL, to $S_{\rm max}-1$. In the case of models for which this second
eigenvalue can be computed, such as translation invariant models on a lattice, this is again
consistent with a well-known fact, namely that the lowest excitation are simple spinwaves.
But in the case of arbitrary coupling constants and spin magnitudes in one dimension
it proves that the first excited state is represented in the subspace of  ``one overturned
spin" (with respect to the fully polarized ground state), i.e., $S^3=S^3_{\rm max} -1$,
which is a new result.

More generally, the FOEL property can help with determining the spectrum of 
the Heisenberg model at low energies, whether by numerical or other means,
in the following way. Suppose $H$ is a Hamiltonian with the FOEL property.
Diagonalize $H$ in the subspaces $\H^{(S_{\rm max}-n)}$, for 
$n=0,1,\ldots N$, and select those eigenvalues that are less or equal than 
$E(H,S_{\rm max}-N)$.
It is easy to see that the FOEL property implies that this way you have obtained
{\em all} eigenvalues of the full $H$ below $\leq E(H,S_{\rm max}-N)$.

This is interesting because you only had to diagonalize the Hamiltonian in an invariant subspace 
that is explicitly known (by the representation theory of $SU(2)$) and of relatively low dimension:
$\dim(\H^{(S_{\rm max}-n)})$ is $O(L^{n})$, while the full Hilbert space has dimension $(2J+1)^L$ for $L$ spin $J$ variables.

\subsection{The ground states of fixed magnetization for the XXZ chain}\label{sec:droplet}

The spin 1/2 XXZ  ferromagnetic chain with suitable boundary conditions, or defined on the 
appropriate infinite-chain Hilbert space has low-energy states that can be interpreted as
well-defined magnetic domains in a background of opposite magnetization \cite{NS_droplet,Ken}.
Using the techniques we have used for proving FOEL, we can rigorously determine the
dispersion relation of a finite droplet of arbitrary size.

The spin 1/2 XXZ chain can, in principle,  be diagonalized using the Bethe Ansatz
\cite{KBI}. There are two complications that may prevent one from obtaining the desired
information about its spectrum. The first is that a complete proof of completeness of the
Bethe Ansatz eigenstates has been obtained and published only for the XXX chain ($q=\Delta=1$),
although the corresponding result for the XXZ chain has been announced quite some time
ago \cite{Gut}. The second problem is that the eigenvalues are the solutions of complicated
sets of equations, such that proving statements as the one we discuss here, may be
very hard.

For brevity, let us consider the XXZ Hamiltonian for the inifinite chain defined on 
the Hilbert space generated by the orthonormal set of vectors representing $n$ down
spins in an infinite ``sea'' of up spins, and let us denote this space by $\H_n$. Define
$$
E(n)=\inf \spec (H\vert_{\H_n}).
$$
As before, the relation between $q\in(0,1)$ and the anisotropy parameter $\Delta$
in the XXZ Hamiltonian \eq{hamXXZ} is given by $\Delta=(q+q^{-1})/2$.

\begin{theorem}
For $n\geq 1$, we have
$$
E(n)=\frac{(1-q^2)(1-q^n)}{(1+q^2)(1+q^n)}.
$$
Moreover, $E(n)$ belongs to the continuous spectrum and is the bottom of a band of 
width
$$
4q^n\frac{1-q^2}{(1+q^n)(1-q^n)}.
$$
\end{theorem}

The states corresponding to this band can be interpreted as a droplet of size $n$ 
with a definite momentum. The formula for the width indicates that the ``mass'' of a
droplet diverges as $n\to \infty$. The proof of this result will appear in a separate
paper \cite{NSS_inprep}. If one looks back at the finite-volume eigenvalues
$E(H_k, k/2-n)$ that converge to the bottom of the band in the infinite-volume
limit ($k\to\infty$, $n$ fixed), the property (iv) amounts to property that the ground
state energy of a droplet of fixed size $n$ is strictly monotone decreasing in the
volume. Moreover the finite-volume eigenvalues can be related to E(n), in the
above limit, by using the generalization of the Perron-Frobenius result
stated in Theorem \ref{thm:PF}.
 
\subsection{Aldous' Conjecture for the Symmetric Simple Exclusion Process}

The Symmetric Simple Exclusion Process (SSEP) is a Markov process defined
on particle configurations on a finite graph $\Lambda$. For our purposes it is 
convenient to define the process as a semigroup on 
$\H_\Lambda \cong l^2(\Omega_\Lambda)$, where $\Omega_\Lambda$ is the space
of configurations $\eta:\Lambda\to\{0,1\}$. One thinks of $\eta(x)=1$ to indicate the
presence of a particle at the vertex $x$. Let $L$ be defined in $\H_\Lambda$
by the formula
\be
(Lf)(\eta) = \sum_{x\sim y}\sum_\eta J_{xy}(f(\eta)-f(\eta^{xy}))
\label{defL}\ee
where $\eta^{xy}$ denotes the configuration obtained form $\eta$ by interchanging the
values of $\eta(x)$ and $\eta(y)$. The parameters $J_{xy}$ are positive numbers
representing the jump rate at the edge $x\sim y\in\Lambda$. 

Clearly, the number of particles is a conserved quantity of the process. Concretely,
this means that $H_\Lambda$ decomposes into a direct sum of invariant subspaces
$H_\Lambda^{(n)}$, $n=0,\ldots,\vert\Lambda\vert$, where  $H_\Lambda^{(n)}$
consists of all functions supported on configurations $\eta$ that have extactly $n$ particles,
i.e., $\sum_x \eta(x) = n$. $0$ is a simple eigenvalue of each of the restrictions
$L\vert_{H_\Lambda^{(n)}}$, and $L$ is non-negative definite. For each $n$, the corresponding invariant measure is the uniform distribution on $n$-particle configurations.

Let $\lambda(n)$ denote the smallest positive eigenvalue of $L\vert_{H_\Lambda^{(n)}}$. 
Since the dynamics of the SSEP is given by the semigroup $\{e^{-tL}\}_{t\geq 0}$, 
$\lambda(n)$, determines the speed of relaxation to the invariant measure.

The following conjecture is known as Aldous' Conjecture but his website \cite{Ald}
Aldous states that it arose in a conversation with Diaconis. So, maybe it should be
called the Aldous-Diaconis Conjecture.

\begin{conjecture}\label{con:aldous}
$$
\lambda(n)=\lambda(1), \quad \mbox{ for all } 1\leq n\leq \vert\Lambda\vert-1.
$$
\end{conjecture}

Apart from being a striking property, namely that the relaxation rate should
be independent of the number of particles, it could also be very useful. The SSEP
for one particle is just a random walk on the graph $\Lambda$, and many 
powerful techniques are available to study the relaxation rate of random walks.
In physical terms one would say that the conjectured property reduces the many-body problem
of finding the relaxation rate for $n$ particles to a single one-particle problem.  

\begin{proposition}
If the ferromagnetic spin-$1/2$ Heisenberg model with coupling constants $J_{xy}$ 
on a graph $\Lambda$ satisfies FOEL, then Conjecture \ref{con:aldous} holds for the SSEP 
on $\Lambda$ with jump rates $J_{xy}/2$.
\end{proposition}
\begin{proof}
The proof is based on the unitary equivalence of $L$ and the ferromagnetic spin $1/2$ 
Heisenberg Hamiltonian $H$ given by
$$
H = \sum_{x\sim y\in\Lambda} J_{xy}\left(\frac{1}{4}\idty - \vec{S}_x\cdot\vec{S}_y\right).
$$
The unitary transformation
$U:L^2(\Omega_\Lambda)\to\H_\Lambda=(\Cx^2)^{\otimes\vert\Lambda\vert}$,
that relates $L$ and $H$ is explicitly given by
$$
L^2(\Omega_\Lambda)\ni f\mapsto Uf
=\psi=\sum_\eta f(\eta)\ket{\eta}, \quad \mbox{where } S^3_x\ket{\eta}=(\eta_x-1/2)\ket{\eta}
$$
To see this note that
$$
1/4-\vec{S}_x\cdot\vec{S}_y=(1-t_{xy})/2,
$$
where $t_{xy}$ interchanges the states at $x$ and $y$ in any tensor product vector.
Then
\beann
H\psi&=&\frac{1}{2}\sum_{x\sim y} \sum_\eta f(\eta)J_{xy}(1-t_{xy})\ket{\eta}\\
&=&\frac{1}{2}\sum_{x\sim y}\sum_\eta J_{xy}(f(\eta)-f(\eta^{xy}))\ket{\eta}\\
&=&\frac{1}{2}\sum_\eta (Lf)(\eta)\ket{\eta}\, .
\eeann
Therefore, $HUf=ULf$, for all $f\in L^2(\Omega_\Lambda$.

Under this unitary transformation, the particle number becomes the third component
of the total spin:
$$
S^3_{\rm tot} = -\vert\Lambda\vert/2 +n.
$$ 
The unique invariant measure of SSEP for $n$ particles is the
uniform measure on $\{\eta\in\Omega_\Lambda \mid \sum_x \eta_x=n\}$.
The corresponding state for the spin model belongs to the unique multiplet of 
maximal total spin, i.e., is a ground state. $\lambda(n)$ is the next eigenvalue of
$H$ is the same value of total $S^3$. Since the first excited state of $H$, by FOEL,
is a multiplet of total spin $S_{\rm max}-1$, this eigenvalue has an eigenvalue
with any value of $S^3$ in the range $-S_{\rm max}+1,\dots,S_{\rm max}-1$. 
We have $S_{\rm max}=\vert\Lambda\vert/2$. Therefore, this corresponds to the range
$1\leq n\leq \vert\Lambda\vert -1$. Hence, $\lambda(n)$ is independent of $n$ in
this range.
\end{proof}

In combination with Theorem \ref{thm:main}, this proposition has the following 
corollary.

\begin{corollary}
Conjecture \ref{con:aldous} holds for chains.
\end{corollary}

Our partial result for trees (not discussed here) also implies Conjecture
\ref{con:aldous} for arbitrary finite trees as well as some graphs derived from trees. 
These cases of the Aldous-Diaconis conjecture were previously know \cite{Bac,HJ}, 
as well as some other examples where one can compute $\lambda(n)$ exactly 
\cite{DS-C,FOW,DS}. Needless to say, a full proof of FOEL, the Aldous-Diaconis Conjecture, 
or even a proof for additionial special cases, would be of great interest.
An interesting direction for generalization considered by Aldous
is to also establish the analogous formula for the spectral gap for a
card-shuffling model with full $SU(n)$ symmetry, which restricts to the
SSEP when one considers cards of only two colors. 

\acknowledgement{This work was supported in part by the
National Science Foundation under Grant \# DMS-0303316.
B.N. also thanks the Erwin Schr\"odinger Institute, Vienna, where part of this work was done, 
for financial support and the warm and efficient hospitality it offers.

\providecommand{\bysame}{\leavevmode\hbox to3em{\hrulefill}\thinspace}

\end{document}